\documentclass[preprint]{aastex}

\slugcomment{Submitted to the Astrophysical Journal Letters.}

\shorttitle{Blue Quasars and  Blazar Unification Schemes}
\shortauthors{Georganopoulos}

\begin{document}

\title{Blue Quasars and Blazar Unification Schemes}

\author{Markos Georganopoulos}
\affil{Max Planck Institut f\"{u}r Kernphysik, Postfach 10 39 80,  Heidelberg, D 69029, Germany}
\email{markos@mickey.mpi-hd.mpg.de}

\begin{abstract}

Blue quasars (BQs) are sources  with  
strong broad emission lines and flat hard X-ray spectra, properties
that resemble  classical flat spectrum radio quasars (FSRQs), and 
high peak frequencies and  steep soft X-ray spectra, properties
that resemble 
intermediate or high peak frequency
 BL Lacertae objects
(IBLs and HBLs respectively).
BQs challenge our understanding of  blazar properties
in terms of a luminosity sequence, which  makes their incorporation into
current blazar unification schemes  problematic. In this work we show that
this situation can be remedied if, in addition to the intrinsic luminosity, 
the orientation of the blazar jet is explicitly considered. We show, 
using published data, that the recently studied BQs are relatively misaligned
blazars, and we examine the predicted aligned population.
We examine both possible cases,
sources with  pure  synchrotron 
spectra and sources with an optical--UV thermal contribution, 
for both constant
velocity and accelerating jets. We show that the aligned sources 
are similar to  FSRQs, and we suggest ways to distinguish
between constant velocity and accelerating flows.
We point out that IBLs are more aligned and less powerful than BQs and 
we address the very different emission line properties of 
these  sources which display 
similar spectral energy distributions.
\end{abstract}


\section{INTRODUCTION}
Blazar--type active galactic nuclei (AGNs) are
characterized by a luminous and rapidly  variable spectral energy distribution
(SED) extending from radio up to GeV and TeV energies \citep{urry95}.   
The blazar SED is characterized by two  components.
The first one  peaks at IR to X-ray energies and it is most probably
 synchrotron radiation from  electrons in a relativistic jet pointing close to
the line of sight. The second one peaks at 
GeV-TeV energies and, according to leptonic models (for a recent review see
B\"ottcher 1999),  is  inverse Compton (IC)
emission from the same electron population, synchrotron self Compton (SSC) 
scattering 
in the case of  synchrotron seed photons 
and external Compton (EC) scattering  in the case of 
external seed photons.

Unified schemes  attempt to understand the properties of
blazar samples using physical models coupled to  scaling and/or geometrical
arguments. 
One of the first unification problems was  the differences
between two  types of  BL Lacertae objects (BLs, blazars 
with  weak broad 
emission lines), the high peak frequency   BLs
(HBLs) and the low peak frequency BLs (LBLs).
\citet{maraschi86} proposed that these differences 
could be explained assuming that the jet flow is accelerating and that 
 LBLs  form a smaller  angle  between the jet axis and line of sight 
than  HBLs.
\citet{sambruna96} argued that the range of observed properties
    in blazars cannot 
be reproduced solely under the different   orientation hypothesis, 
and that physical 
changes need to be invoked to explain the gradual change of observed properties
going from flat spectrum radio quasars (FSRQs) to LBLs to HBLs.
 The gap between HBLs and LBLs was filled with the discovery 
\citep{laurent99} of the  intermediate  BLs (IBLs), sources  with peak
frequencies at optical-UV energies and with 
 properties intermediate between those of HLBs and LBLs.
\citet{georganopoulos98} proposed a  unification scheme for BLs
based on two parameters, the intrinsic
luminosity of the source and  the orientation of the jet relative
to the observer.

\citet{fossati98} and \citet{ghisellini98}
pointed out a sequence of blazar properties as a function of source power.
 As the source power  increases, the emission line luminosity and 
  the ratio of Compton to synchrotron luminosity
are increasing, while the synchrotron 
peak frequency $\nu_{s}$ and the IC peak frequency decrease.  
Recent multiwavelength studies support this scheme (e.g. \citet{kubo98}).  
Under this scheme, sources with strong emission lines should  have low
$\nu_{s}$  and IC--dominated flat X-ray spectra, and sources with high
$\nu_{s}$  should be lineless and have steep X-ray spectra dominated
by the high energy tail of the synchrotron emission.   
Recent observations (\citet{sambruna97}; \citet{perlman98}; \citet{sambruna00})
  revealed a population of high energy peaked
FSRQs, dubbed blue quasars (BQs),
with properties that challenge  the above scheme. These   are sources with {\it ROSAT} (0.1--2.4 KeV) 
spectra steeper than those of  FSRQs,
similar to those of HBLs.
At the same time they exhibit strong emission lines and, 
those observed with  {\it ASCA}  (Sambruna et al. 2000), 
flat  hard X--ray (2--10 KeV) spectra similar to classical FSRQs. 
Sambruna et al. (2000) presented SEDs and {\it ASCA} observations of 4 BQs.
 As they note, these sources are
peaking at optical--UV energies (although with the current data the location of the SED peak
is still somewhat uncertain), and   it is not clear if the optical to UV emission is 
synchrotron as in BLs  or if it has a thermal
emission component as  in sources like 
3C 273. They also point out that while these BQs have similar SEDs  and $\it ASCA$ spectral indices
 ($\alpha_{2-10\; KeV}\approx0.8$) to   IBLs,
they have much stronger broad emission lines.
In this work we address  the
unusual properties of BQs and their implications on 
 blazar unification.

\section{BLUE QUASARS}

In Table \ref{tbl1} we present  published data for the 4 BQs
of Sambruna et al. (2000) and for two IBLs with sufficient observations,
which we use for comparison
purposes: 
 source name,  source type, 
 ratio $R$ of the extended (nonbeamed) to core (beamed) radio power, 
which is a measure  of the alignment between the jet axis 
and the line of sight, 
  extended radio power $P_{ext}$, and 
 broad line region (BLR) power $L_{BLR}$.
The smaller $R$  and the higher $P_{ext}$ and $L_{BLR}$ of BQs indicate
that BQs are more powerful and less aligned that  IBLs.
Additional evidence for the relative misalignment of 0923+392 
\citep{kollgaard90} and 0405-123 \citep{morganti93}
comes  from the FR II like radio morphology of these sources. 
Given the relative misalignment of BQs we ask 
how the SED of the aligned version of these
sources  looks, and how these aligned sources compare  with sources
found in current blazar samples.    
We examine   both cases for  the optical--UV flux being only synchrotron
or having an additional thermal contribution from an accretion disc.
We use the  jet formalism of \citet{kirk97}, 
modified to include IC losses due to an external photon
field and an  angle dependent emission for both a constant Lorentz factor
and an accelerating jet \citep{georganopoulos98}. 
In this work the IC emission is not modeled. Modeling the hard X-ray emission,
which is  probably  due to SSC  \citep{kubo98}, is highly complicated for
the inhomogeneous jets studied here. Although we cannot   
address the hard X-ray emission  quantitatively, we can discuss qualitatively 
the scaling of the observed hard X-ray flux of the BQs as a 
function of beaming, since the beaming behavior of the IC component 
is well understood \citep{dermer95}.

\subsection{Case A:  Thermal component}

\begin{figure}
\epsscale{0.8}
\plotone{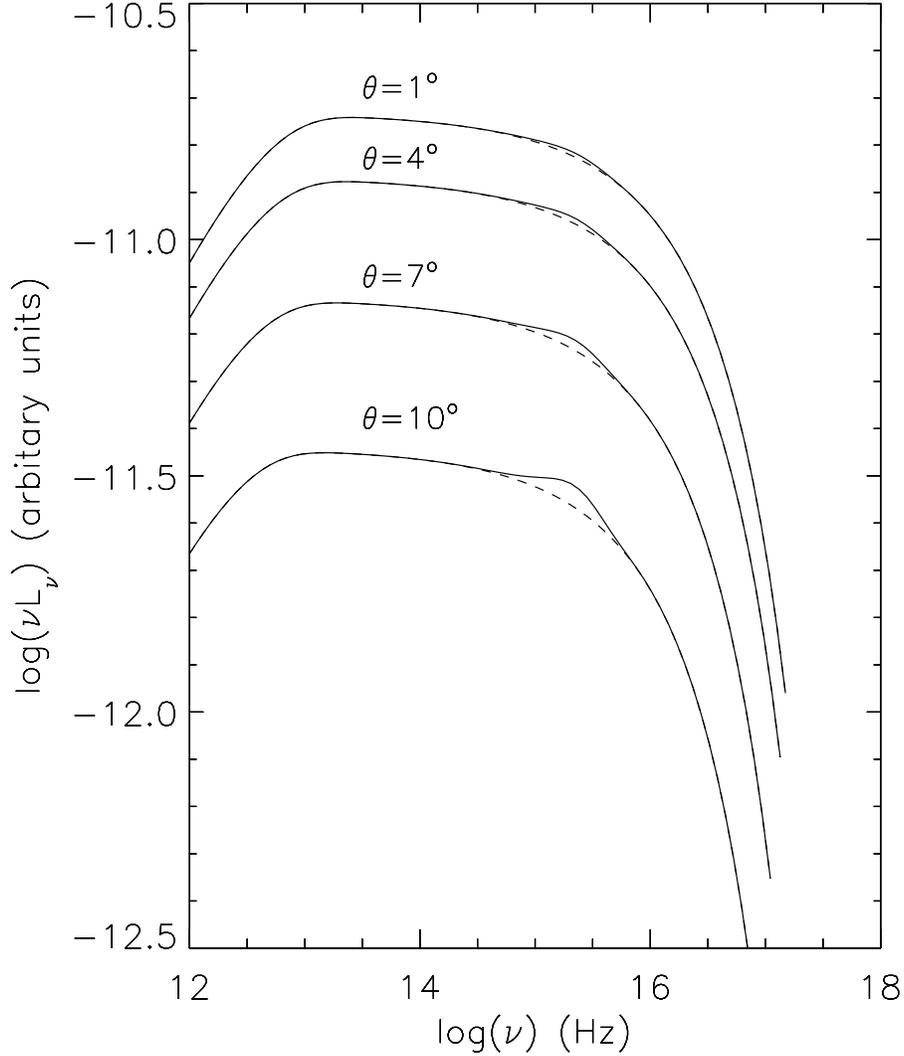}
\caption{The SED of  a constant Lorentz factor jet for a 
range of angles 
$\theta=1^{\circ},4^{\circ},7^{\circ},10^{\circ}$ between the jet axis and
the line of sight. 
The solid line corresponds to the total emission, 
including the BB  emission, 
while the broken line corresponds to the synchrotron emission
only. We assume here, as in all the models that include a BB,  that the BB 
luminosity is equal to the kinetic power in the jet.
Model parameters: Bulk motion Lorentz factor $\Gamma=5$, jet cross section
radius $R=10^{16}$ cm, magnetic field $B=1.0$ G, maximum electron energy
$\gamma_{max}=5 \; 10^4$, external photon energy density $U_{ext}=0.1\; U_{B}$,
where $U_{B}=B^2/8\pi$ is the magnetic field energy density, 
jet length $Z=2\; 10^{17}$ cm.}
\label{fig1}
\end{figure}

We study first  a constant Lorentz factor $\Gamma$ jet together with a
thermal component modeled as black body (BB)
radiation.  In Figure \ref{fig1} we plot the SED for a range of 
angles $\theta$ between the jet axis and the line of sight. 
The jet emission is strongly affected by Doppler boosting
and,  as  $\theta$  decreases, the non--thermal
SED shifts mostly upward to higher apparent luminosities with a 
slight shift to higher peak 
frequencies, since $L\propto \delta^{3+\alpha} L_{0}$
and $\nu\propto \delta \nu_{0}$, where  $\alpha$ is the spectral index,
 $\delta$ is the usual Doppler factor
$\delta=1/(\Gamma(1-\beta_{\Gamma}\cos\theta))$, 
and the subscript $0$ refers to
 quantities in the flow comoving frame. Since the BB and the BLR emission are
 not  a function of $\theta$, the relative 
contribution of the thermal 
component and the equivalent width (EW) 
of the emission lines (which are  not plotted here, but are  assumed 
to have a fraction of the BB luminosity)
are  reduced
as $\theta$  decreases. 
We expect the IC 
{\it ASCA}  component to either follow the increase of the synchrotron one 
if it is due to SSC emission or to increase even faster if it is due to EC 
emission, since $L_{SSC}\propto
\delta^{3+\alpha}$ and $L_{EC} \propto \delta^{4+2\alpha}$  \citep{dermer95}.  
In both cases, and depending on $L_{BLR}$, the aligned source will
look like an LBL or a classical FSRQ, possibly similar to 3C 279, 
a source showing evidence of a thermal component in its optical--UV
spectrum \citep{pian99}. 
It is interesting to note that for 3C 279, $\log L_{BLR}=44.76$ erg s$^{-1}$
 \citep{cao99}, 
in the $L_{BLR}$  range  of the BQs examined here.

\begin{figure}
\epsscale{0.8}
\plotone{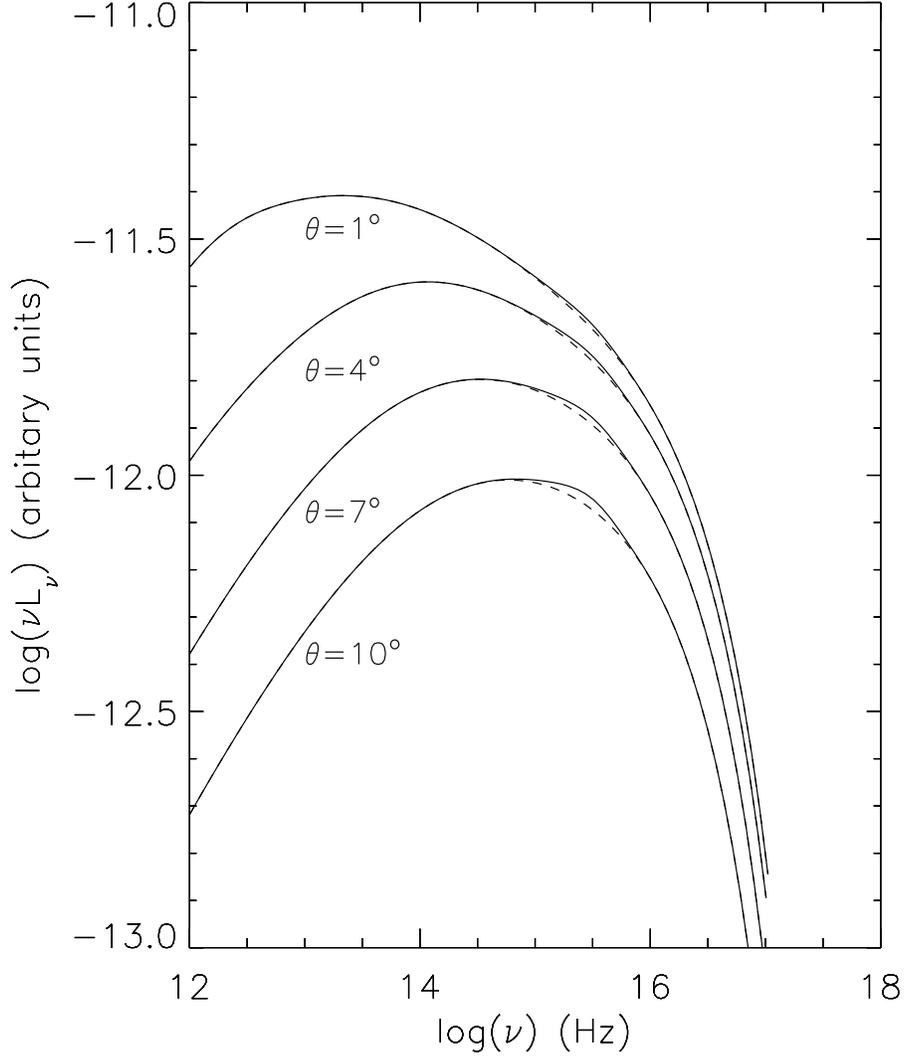}
\caption{The SED for an accelerating  jet. Description 
as in Figure \ref{fig1}. The bulk motion Lorentz factor 
$\Gamma$ increases along the jet: $\Gamma= \Gamma_0 (z/z_0)^{1/2}$.
The jet has a parabolic form, and 
its  radius  is $R=R_0(z/z_0)^{1/2}$, where $\Gamma_0=3$, 
$R_0=z_0=10^{16}$ cm. The magnetic field decays as $1/R$, 
$B=B_0\;(z/z_0)^{(1/2)}$, $B_0=1.0$ G. The rest of the
parameters are the same as for  the constant Lorentz factor model.}
\label{fig2} 
\end{figure}

We study now the case of an accelerating jet.  Higher
frequencies emerge closer to the base of the jet, are  
characterized by  lower  $\Gamma$, and are therefore less sensitive
to angle variations compared to the lower frequencies
 \citep{georganopoulos98}. 
As   can be seen from the solid lines in Figure \ref{fig2},
 as $\theta$  decreases, the non--thermal
SED shifts  upward to higher apparent luminosities, but this time with a  
large  shift to lower $\nu_{s}$, since the higher $\Gamma$
of the low--frequency emitting plasma results to a larger increase in the
low--frequency  boosting. 
The relative 
contribution of the thermal 
component and the EW 
of the emission lines drop  as the source becomes more  aligned, 
although this is 
more gradual  compared
to the apparent luminosity and $\nu_{s}$ changes observed due to the
 frequency--dependent 
Doppler boosting. 
The IC {\it ASCA}  component must be   mostly  due to the same plasma 
responsible for the sub-mm--IR emission, and is  characterized by a 
$\Gamma$ higher than the $\Gamma$  of the X--ray emitting
plasma. Therefore, as  the source becomes more aligned, the IC component
will progressively dominate over the synchrotron at lower  frequencies.
Qualitatively, the aligned population in this case is
similar to the one predicted for a constant $\Gamma$ jet.
A  discriminator  between the two cases could be the detailed
form of the SED
for low $R$ (more misaligned) objects. One would expect the synchrotron 
and the BB peak frequencies
of the SED to be further apart in the constant $\Gamma$ 
 case, since in  this case the synchrotron peak frequency has to be
$\nu_{s}\sim 10^{12.5-13.5}$ Hz, close to the   synchrotron peak frequency
of classical FSRQs and LBLs, and the BB peak frequency $\sim 10^{15}$ Hz.

\subsection{Case B: Synchrotron emission}

\begin{figure}
\epsscale{0.8}
\plotone{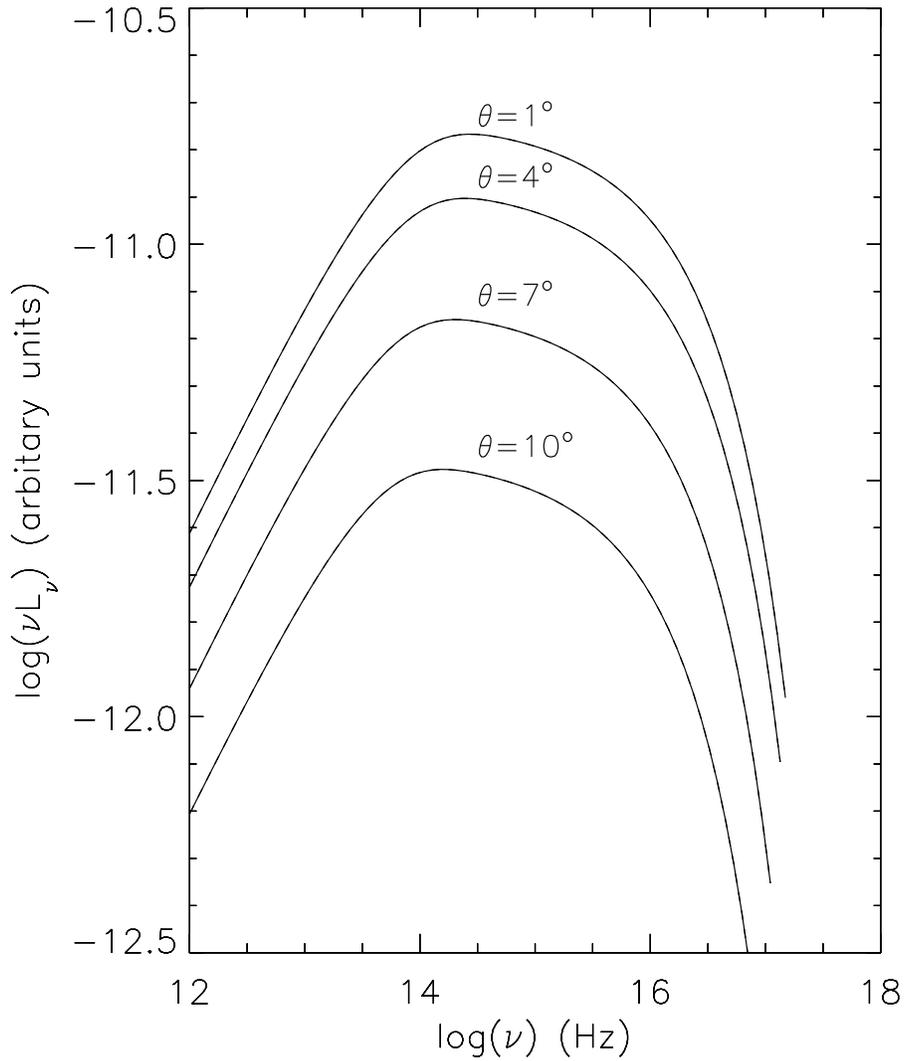}
\caption{The SED of  a constant Lorentz factor jet 
without a BB component for a range of angles 
$\theta=1^{\circ},4^{\circ},7^{\circ},10^{\circ}$ between the jet axis and
the line of sight. All the parameters are the same as in Figure \ref{fig1},
except that the external photon energy density is zero, and that the length
of the jet is shorter, $Z=5\; 10^{16}$ cm.}
\label{fig3}
\end{figure}

We now assume that the observed SED of the BQs, with a peak at 
$\nu_{s}\sim 10^{14-15}$, is   synchrotron radiation with  no 
significant  thermal  contribution.
As can be seen in Figure \ref{fig3}, aligning a constant $\Gamma$ 
jet will boost up the  apparent luminosity and will slightly increase $\nu_{s}$.
The EW of the emission lines   (which are still 
assumed to be powered by an accretion disc) will be reduced,
 but the IC component will
be boosted either as much as the synchrotron component (SSC case) or  more
(EC case). In both cases the aligned sources will be 
bright sources with $\nu_{s}\approx10^{14-15}$, low  EW emission lines,
and  a  spectral flattening between {\it ROSAT} and  {\it ASCA} 
 energies, due to IC emission. 
Such high luminosity, high peak frequency sources,
 with  hard X--ray IC   spectra flatter  than the synchrotron soft X--ray 
spectra
have not been observed,
and there is no obvious selection effect acting against detecting them.
If such sources existed they would have been observed in X--ray selected
BL samples. Therefore, the possibility of pure synchrotron emission
and constant $\Gamma$ jets is excluded for BQs.

The case of a synchrotron SED for an accelerating flow 
(Figure \ref{fig2}, broken line) is quite similar 
to the accelerating flow with the BB contribution and it is practically
 impossible  to discriminate between the two without a detailed knowledge of the
SED at optical--UV energies.

\section{DIFFERENCES  WITH IBLS}

 A  question pointed out by Sambruna et al. (2000) is 
how can objects with similar SEDs  like BQs and IBLs   have so different emission line  properties. 
The higher value of $R$  and the smaller $P_{ext}$ and $L_{BLR}$ of the IBLs
we present in Table \ref{tbl1} show that these are less powerful sources
seen under a smaller angle.  
We can qualitatively 'transform' a BQ to an IBL with two  translations, 
one  in angle and 
one in intrinsic  power:  first align the BQ, then decrease its intrinsic
power. The source resulting from the first translation  will have 
 smaller  EW emission lines,
 since decreasing $\theta$ boosts
the synchrotron component, but it will be  more 
luminous  than an IBL. The second step will reduce the luminosity of the 
source down to the IBL luminosity. The similar peak frequencies hold an
interesting clue to the acceleration and cooling processes: 
since the   aligned BQ  will have $\nu_{s}\approx 10^{13}$ Hz 
regardless of an accelerating or constant velocity jet, the decrease of the
intrinsic power must increase  $\nu_{s}$ to typical values for IBLs  
($\nu_{s}\approx 10^{14-15}$ Hz). This means that cooling becomes less 
efficient and/or particles are accelerated to higher energies as 
the intrinsic power of the source is reduced, something that has been 
previously argued by
\cite{ghisellini98}.

We now focus on the {\it ASCA} X--ray emission.
Although the  {\it ASCA} spectra of all the sources presented in Table
\ref{tbl1} are flat ($\alpha_{2-10\; KeV}\approx0.8$),
they differ in the following sense: while in the IBLs the 
hard X-ray luminosity is  $\approx 100$ times 
weaker than the peak luminosity
of the first spectral component \citep{kubo98}, in the BQs this is at most 
10 times weaker.
Given  now the difference in $\theta$, if one was to align the BQs,
this difference would either persist or would become even more pronounced. 
This is because the hard X-ray emission  is either SSC 
($L_{SSC}\propto\delta^{3+\alpha}$) or EC  
($L_{EC} \propto \delta^{4+2\alpha}$)  emission, while the apparent luminosity
of the first component will at most increase as $\delta^{3+\alpha}$
if it is pure synchrotron emission with no contribution from a thermal
component.
If this  difference in the hard X-ray luminosity relative level 
is also present in the peak IC luminosities, then  
the  ratio of the photon to magnetic field energy density
experienced by the emitting particles is higher in high power sources.
Unfortunately, there are no {\it EGRET} detections of the BQs, although
the two IBLs we compare them with have been detected with an IC  luminosity
practically equal to that of the synchrotron component \citep{kubo98}.

\section{DISCUSSION \& CONCLUSIONS}

BQs seem to be relatively misaligned blazars with a relatively de-boosted
synchrotron continuum. 
It is not clear if there is a  thermal component in  the optical--UV spectrum 
of the  BQs (Sambruna et al. 2000).  
If there is,
one can distinguish between an accelerating and constant $\Gamma$ 
flow depending on the peak frequency $\nu_{s}$ of the synchrotron component.
If $\nu_{s}\approx 10^{12.5-13.5}$ Hz, then the flow is most probably
characterized by a single Lorentz factor. Otherwise, if $\nu_{s}$ is close
to the peak of the thermal component, the flow is most probably accelerating.
In both cases the aligned sources will be LBLs or FSRQs, possibly similar
to 3C 279. 
On the other hand, if no significant thermal component can be detected in
BQs, the possibility of a constant $\Gamma$ jet is
excluded, since the aligned version would be a source similar  to  a bright 
HBL,  but with a strong and flat IC hard X-ray spectrum. Such sources are not
observed, and there are no obvious selection criteria against their 
detection. Clarification therefore of the nature of the SED 
can  provide valuable insight on  the flow characteristics. 

Orientation aside, sources of different intrinsic power are 
different. The synchrotron frequency
$\nu_{s}$ is lower for higher power sources.
This suggests that cooling is stronger
in high power
sources. 
In addition, the relative level of the hard X-ray IC luminosity hints that
 the ratio of the comoving photon  
to magnetic field energy density is higher 
in high power
sources. Measurements of the IC peak luminosity are needed though 
to check if this is really the case.

 The fact that  cooling is
 not invariant under an
increase of the intrinsic power suggests that powerful sources are not
simply  scaled up versions of weak sources. 
Since both the orientation and the intrinsic power strongly
affect the observed characteristics of a source, 
we are  naturally led to
a two-dimensional unification scheme for blazars. In this scheme,
the BQs observed by Sambruna et al. (2000) appear to be 
 intrinsically powerful  sources, whose jet  angle to the line of sight
is  relatively larger than that of typical blazars.

\acknowledgments This work  was supported by  the European Union
TMR programme under contract FMRX-CT98-0168.

\clearpage

\begin{deluxetable}{lcccc}
\tablecolumns{7}
\tabletypesize{\scriptsize}
\tablecaption{Observational data. \label{tbl1}}
\tablewidth{0pt}
\tablehead{
\colhead{Source} & \colhead{Type} & R &\colhead{$\log P_{ext}$} &
 \colhead{$\log L_{BLR}$} \\
	&	&  &(W Hz$^{-1}$) &(erg s$^{-1}$) } 
\startdata
\objectname[]{0405-123} & BQ &0.57 \tablenotemark{d} &27.4 \tablenotemark{d}  & 46.02 \tablenotemark{a}   		\\
\objectname[]{0736+017} & BQ & 3.42 \tablenotemark{d} & 25.9 \tablenotemark{d}& 44.43 \tablenotemark{a}			\\
\objectname[]{0923+392} & BQ &0.44 \tablenotemark{e} &26.85 \tablenotemark{e}&  45.88  \tablenotemark{a}				\\
\objectname[]{1150+497} & BQ &0.39 \tablenotemark{e} & 26.78  \tablenotemark{e}  & 44.56  \tablenotemark{a}		\\ \tableline
\objectname[]{0235+164} & IBL&61.4 \tablenotemark{b}&25.7 \tablenotemark{b} & 43.92  \tablenotemark{a} \\
\objectname[]{0735+178} & IBL&195 \tablenotemark{c}&25.0 \tablenotemark{c} &  ---	
\enddata

\tablenotetext{a}{ Cao \& Jiang 1999}
\tablenotetext{b}{ Kollgaard et al. 1996}
\tablenotetext{c}{ Perlman \& Stocke 1994}
\tablenotetext{d} { Morganti et al. 1997}
\tablenotetext{e} { Hooimeyeret al. 1992}
\end{deluxetable}
\clearpage

\clearpage
\clearpage

\clearpage

\end{document}